\documentclass{article}

\usepackage{spconf,amsmath,graphicx}
\usepackage{amssymb}
\usepackage{graphics} % for pdf, bitmapped graphics files
\usepackage{booktabs}
\usepackage{multirow}
\usepackage{url}
\usepackage{hyperref}
\usepackage{cleveref}

\newcommand{\mysection}[1]{\vspace*{-2mm}\section{#1}\vspace*{-2mm}}
\newcommand{\mysubsection}[1]{\vspace*{-1mm}\subsection{#1}\vspace*{-1mm}}

\crefformat{footnote}{#2\footnotemark[#1]#3}

\title{GANStrument: Adversarial Instrument Sound Synthesis\\ with Pitch-invariant Instance Conditioning}
\name{Gaku Narita, Junichi Shimizu, and Taketo Akama}
\address{Sony Computer Science Laboratories, Tokyo, Japan}
\begin{document}
%\ninept
%
\maketitle
\begin{abstract}\vspace*{-1mm}
We propose GANStrument, a generative adversarial model for instrument sound synthesis.
Given a one-shot sound as input, it is able to generate pitched instrument sounds that reflect the timbre of the input within an interactive time.
By exploiting instance conditioning, GANStrument achieves better fidelity and diversity of synthesized sounds and generalization ability to various inputs.
In addition, we introduce an adversarial training scheme for a pitch-invariant feature extractor that significantly improves the pitch accuracy and timbre consistency.
Experimental results show that GANStrument outperforms strong baselines that do not use instance conditioning in terms of generation quality and input editability.
Qualitative examples are available online\footnote{\label{1}Audio examples are available on \url{https://ganstrument.github.io/ganstrument-demo/}}.
\end{abstract}

\mysection{Introduction}
\label{sec:introduction}

Since the advent of computers, many musicians and researchers have explored ways to generate music with computers. 
There are two main approaches: direct synthesis of music sounds including melody and accompaniment, and single note synthesis followed by playing symbolic music like MIDI.
The former enables end-to-end music synthesis, but has low controllability of generation.
The latter enables MIDI and timbre to be independently controlled and is compatible with production flows in the music industry.
In this paper, we tackled instrument sound synthesis for the latter approach.

Realistic instrument sounds are typically synthesized with samplers that utilize recorded one-shot sounds.
Although arbitrary sound materials can be exploited, it is difficult to synthesize a completely new timbre or intelligently combine multiple sounds.
In contrast, recently reported deep generative models for audio synthesis \cite{engel2017neural,engel2018gansynth,DBLP:conf/ismir/LuoAH19,nistal2020drumgan,engel2020ddsp,shan2022differentiable} have the potential to generate and mix a variety of timbres by exploring the latent space.
Our aim is to design a neural synthesizer that combines the flexibility of samplers with the generative power of deep networks, thereby enabling users to freely control the timbre by leveraging existing sound materials.
For practical use, it needs to not only be generalized to a variety of inputs but also be able to generate high-quality audio with accurate pitch within an interactive time.

Towards this end, we present GANStrument, a novel neural instrument sound synthesizer.
To enable the model to accept various inputs, we focus on instance conditioning \cite{casanova2021instanceconditioned}, a new GAN training scheme for conditioning a model on input features.
In addition, we present a pitch-invariant feature extractor based on adversarial training that disentangles the latent space and significantly improves pitch accuracy and timbre consistency.
Furthermore, use of the modern GAN architecture, parallel sampling with spectrogram representation, and carefully designed audio inversion enables high-quality audio to be generated within an interactive time.

\mysection{Method}
\label{sec:method}

\begin{figure*}[t]
  \centering
     \includegraphics[height=3.5cm]{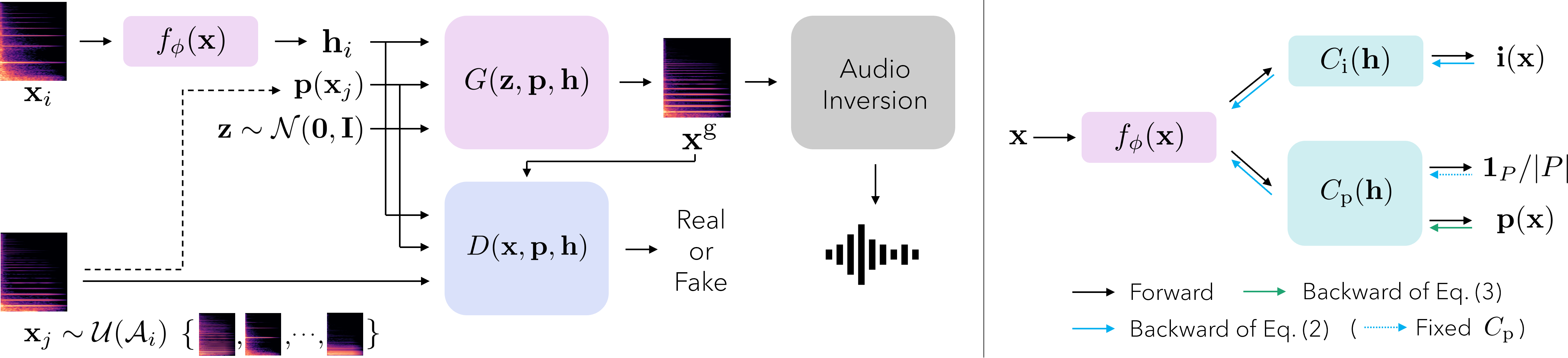}
     \vspace*{-4mm}
     \caption{Overview of GANStrument. Left side shows training and inference pipeline of generative model. Right side depicts adversarial training scheme of feature extractor.}
     \label{fig:model_overview}
\end{figure*}

As depicted in Fig. \ref{fig:model_overview}, the input waveform is first transformed into a mel-spectrogram $\mathbf{x}_i$, and its feature $\mathbf{h}_i$ is extracted with the feature extractor $f_\phi$.
It is fed into the generator $G$ together with pitch $\mathbf{p}$ and noise $\mathbf{z}$ to synthesize a mel-spectrogram $\mathbf{x}^\mathrm{g}$, which is transformed into a waveform by optimization-based audio inversion.
Feature extractor $f_\phi$ is first trained by incorporating the pitch-adversarial loss into a standard classification loss.
Using a frozen feature extractor, we jointly train the generator $G$ and discriminator $D$ with input neighborhoods as real samples.

\mysubsection{Instance conditioning}
\label{subsec:instance_conditioning}

Class-conditional GANs partition the entire data distribution into multiple distributions without overlap.
In contrast, instance-conditioned GAN \cite{casanova2021instanceconditioned} partitions the entire data distribution into many overlapping local distributions and thereby model a complex distribution.
By conditioning both the generator and discriminator with instance feature $\mathbf{h}_i = f_\phi(\mathbf{x}_i)$ and pitch $\mathbf{p}$, we model a local distribution of instance neighborhood $p(\mathbf{x} | \mathbf{h}_i, \mathbf{p})$ and represent the entire data distribution $p(\mathbf{x})$ as a mixture of these distributions: $\sum_{\mathbf{h}_i} \sum_{\mathbf{p}} p(\mathbf{x} | \mathbf{h}_i, \mathbf{p}) p(\mathbf{p} | \mathbf{h}_i) p(\mathbf{h}_i)$.

We follow the training procedure of Casanova et al. \cite{casanova2021instanceconditioned}.
For input $\mathbf{x}_i$, let $\mathcal{A}_i$ be the $L_2$-based $k$ nearest neighbors of $\mathbf{x}_i$ over the feature space defined by $f_\phi$.
As shown in Fig. \ref{fig:model_overview}, we sample neighborhood data point $\mathbf{x}_j$ from the uniform distribution $\mathcal{U}(\mathcal{A}_i)$.
Then $\mathbf{x}_j$ is used as a real sample together with a generated sample $\mathbf{x}^\mathrm{g}$ to train the discriminator $D$, and its corresponding pitch $\mathbf{p}(\mathbf{x}_j)$ is fed into both the generator $G$ and discriminator $D$ for conditioning.
Formally, we jointly optimize $G$ and $D$ using the following min-max game:
\begin{eqnarray}
  \label{eq:gan_loss}
  \min_G \max_D \mathbb{E}_{\mathbf{x}_i \sim p(\mathbf{x}), \mathbf{x}_j \sim \mathcal{U}(\mathcal{A}_i)} [\log D(\mathbf{x}_j, \mathbf{p}(\mathbf{x}_j), \mathbf{h}_i)] + \nonumber \\
  \mathbb{E}_{\mathbf{x}_i \sim p(\mathbf{x}), \mathbf{z} \sim p(\mathbf{z})} [\log (1 - D(G(\mathbf{z}, \mathbf{p}(\mathbf{x}_j), \mathbf{h}_i), \mathbf{p}(\mathbf{x}_j), \mathbf{h}_i))].
\end{eqnarray}

\mysubsection{Pitch-invariant feature extractor}
\label{subsec:feature_extractor}

The simplest way to obtain feature extractor $f_\phi$ is to train a classifier using labeled data.
In our case, for example, we can train an instrument identity classifier on the NSynth dataset \cite{engel2017neural}.
However, features extracted by this classifier contain not only timbre but also pitch information, resulting in a decrease of pitch accuracy, as shown in Sec. \ref{subsec:ablation_study}.
This is because the generator and discriminator confuse the specified pitch $\mathbf{p}$ with remaining pitch information in $\mathbf{h}$, and the training process starts oscillating.

To solve this problem, we propose a pitch-invariant feature extractor based on an adversarial training scheme that enables disentanglement of timbre and pitch in the latent space.
Our training scheme is inspired by previous work such as domain adaptation \cite{ganin2016domain}, image manipulation \cite{lample2017fader}, and music domain transfer \cite{musictranslation}, in which adversarial training was introduced to the bottleneck features.
Let $C_\mathrm{i}(\mathbf{h})$ and $C_\mathrm{p}(\mathbf{h})$ be shallow MLPs that predict instrument identity $\mathbf{i}(\mathbf{x})$ and pitch $\mathbf{p}(\mathbf{x})$, respectively, given a feature $\mathbf{h} = f_\phi(\mathbf{x})$.
As shown on the right side of Fig. \ref{fig:model_overview}, the following objectives are alternately optimized to obtain $f_\phi$:
\begin{eqnarray}
  \label{eq:feature_extractor_loss_fisrt}
  \min_{f_\phi, C_\mathrm{i}} \mathrm{CE}(\mathbf{i}(\mathbf{x}), C_\mathrm{i}(f_\phi(\mathbf{x}))) + \lambda_\mathrm{adv} \mathrm{KL}(\frac{\mathbf{1}_{P}}{|P|} || C_\mathrm{p}(f_\phi(\mathbf{x}))), \\
  \label{eq:feature_extractor_loss_second}
  \min_{C_\mathrm{p}} \mathrm{CE}(\mathbf{p}(\mathbf{x}), C_\mathrm{p}(f_\phi(\mathbf{x}))), \ \ \ \ \ \ \ \ \ \ \ \ \ \ \ \ \ \
\end{eqnarray}
where $|P|$ is the number of pitches, $\mathbf{1}_P \in \mathbb{R}^{|P|}$ is the all-one vector, and $\mathrm{CE}$ and $\mathrm{KL}$ represent cross entropy and Kullback-Leibler divergence, respectively.
The first term of Eq. \eqref{eq:feature_extractor_loss_fisrt} updates $f_\phi$ and $C_\mathrm{i}$ so that instrument identities can be correctly classified, while the second term makes it impossible to classify pitch given $\mathbf{h}$.
Eq. \eqref{eq:feature_extractor_loss_second} updates $C_\mathrm{p}$ to maximize the accuracy of pitch classification given $\mathbf{h}$.
This adversarial training eventually produces instance feature $\mathbf{h}$, which contains little pitch information.
In fact, we re-trained pitch classifier $C_\mathrm{p}$ with the frozen $f_\phi$ using Eq. \eqref{eq:feature_extractor_loss_second}, and the accuracy of pitch classification dropped from 17.4\% to 2.6\% with our adversarial training scheme while that of instrument identity classification remained unchanged at 91.2\%, meaning that feature $\mathbf{h}$ preserves timbre information.

\mysubsection{Audio inversion}
\label{subsec:audio_iversion}

In the field of speech synthesis, learning-based vocoders have achieved high-quality audio synthesis \cite{shen2018natural, kumar2019melgan, kong2020hifi}.
However, several studies \cite{lee2022bigvgan, wu2022ddsp, lee2022adversarial} suggested the difficulty of making neural vocoders generalized to a variety of timbre and pitch, which GANStrument is aimed at generating.
On the other hand, MelNet \cite{vasquez2019melnet} revealed that optimization-based audio inversion can synthesize a variety of audio including music with decent quality by using high-resolution mel-spectrograms (e.g., 256 bins).

Therefore, we use optimization-based audio inversion with high-resolution mel-spectrograms (512 bins).
Mel-spectrogram inversion typically consists of a mel-to-linear frequency-scale conversion and phase restoration using the Griffin-Lim algorithm \cite{griffin1984signal}.
The frequency-scale conversion could be a bottleneck here because a non-negative least-squares problem $\min_{\mathbf{x}_\mathrm{lin}} || \mathbf{F}_\mathrm{mel} \mathbf{x}_\mathrm{lin} - \mathbf{x}_\mathrm{mel} ||^2 \ \mathrm{s.t.} \ \mathbf{x}_\mathrm{lin} \geq 0$ must be solved with the computationally demanding L-BFGS-B algorithm \cite{byrd1995limited}.
In place of L-BFGS-B, $\mathtt{torchaudio}$ \cite{yang2021torchaudio} introduces the first-order gradient method with negative value clipping for faster iteration.
However, it requires a sufficient number of iterations due to random initialization.

We propose a simple yet effective initialization scheme.
First, we solve an unconstrained least-squares problem $\min_{\mathbf{x}_\mathrm{lin}} || \mathbf{F}_\mathrm{mel} \mathbf{x}_\mathrm{lin} - \mathbf{x}_\mathrm{mel} ||^2$ using the divide-and-conquer SVD because the mel filter bank $\mathbf{F}_\mathrm{mel}$ is not well-conditioned.
% which can be computed quickly.
Next, the negative values of the solution are clipped, and the clipped solution is set as the initial value for the iterative method of the first-order gradient method.
This initialization scheme reduces the number of iterations by a factor of 10.

\mysection{Evaluation}
\label{sec:evaluation}

\mysubsection{Experimental setup}
\label{subsec:experimental_setup}

We trained GANStrument on the NSynth dataset \cite{engel2017neural}, a large-scale instrument sound dataset that includes rich annotations such as instrument categories, identities, and pitches.
We extracted 88 pitches (MIDI notes 21–108) and used their first 1-s segments with amplitude normalization and exponential fade-out preprocessing.
For evaluation, we also used single notes of Good-sounds \cite{romani2015real}, with the silence intervals trimmed, and used the same preprocessing as the NSynth dataset.

To compute mel-spectrograms, we used an STFT with a Hann window, a 1024 window size, a 64 hop size, and a 2024 fft size. This was followed by mel-scale conversion with 512 filter banks, resulting in $512 \times 256$ mel-spectrograms.

We utilized the StyleGAN2 \cite{karras2020analyzing} architecture, a state-of-the-art image synthesis model, for the backbone and used a projection discriminator \cite{miyato2018cgans}.
For feature extractor $f_\phi$, we used an architecture that removes the final layer of the discriminator.
We jointly trained the generator and discriminator using the ADAM optimizer with a learning rate of $2.5 \times 10^{-3}$, $(\beta_1, \beta_2)=(0.0, 0.99)$, and $\epsilon=10^{-8}$ for 300k steps with a batch size of $16$.
For training stability, we exploited the training techniques described by Karras et al. \cite{karras2020analyzing} such as $R_1$ regularization and path length regularization. We used $k=50$ for the neighborhood search.

\mysubsection{Generation quality}
\label{subsec:generation_quality}
\begin{table}[t]
  \caption{Generation quality}
  \label{table:generation}
  \begin{tabular}{@{}ccccc@{}}
    \toprule
    \multirow{2}{*}{conditioning}    & \multicolumn{2}{c}{Nsynth (val set)} & \multicolumn{2}{c}{Good-sounds}   \\ \cmidrule(l){2-5} 
                                     & FID$\downarrow$   & Pitch$\uparrow$  & FID$\downarrow$ & Pitch$\uparrow$ \\ \midrule
    pitch                            & 490.1             & 0.831            & 1837.0          & 0.900           \\
    pitch + instrument               & 469.6             & 0.828            & 921.6           & 0.937           \\ \midrule
    \textbf{pitch + instance (ours)} & \textbf{212.3}    & \textbf{0.870}   & \textbf{507.3}  & \textbf{0.946}  \\ \bottomrule
    \end{tabular}
\end{table}

First, we evaluated the fidelity and diversity as well as pitch accuracy of the generated samples.
To validate the proposed approach, we trained two class-conditional GANs as strong baselines: the first model $G_1(\mathbf{z}, \mathbf{p})$ was conditioned on pitch $\mathbf{p}$ and the other $G_2(\mathbf{z}, \mathbf{p}, \mathbf{c})$ was conditioned on both pitch $\mathbf{p}$ and instrument category $\mathbf{c}$ (using 11 NSynth instrument categories).
For fair comparison, we used the same architecture and training parameters for these baselines as in Sec \ref{subsec:experimental_setup}.

To evaluate the Fréchet inception distance (FID) and pitch accuracy, we trained both the instrument category classifier (as an FID feature extractor) and the pitch classifier on the NSynth dataset.
The architecture of these classifiers was a slightly modified version of the discriminator.
They respectively achieved accuracies of 74.3\% and 93.2\% (against the test set).
For evaluation on Good-sounds, we used the same distributions of pitch and category as in the dataset for fair comparison.

Table \ref{table:generation} shows that GANStrument was superior to the baselines on both datasets, suggesting that GANStrument has not only the ability to model the distribution of training data but also the ability of generalization.

\mysubsection{Editability}
\label{subsec:editability_of_inputs}
\begin{table*}[t]
  \caption{Editability}
  \label{table:reconstruction_and_interpolation}
  \centering
  \begin{tabular}{@{}cc|cccc|cccc@{}}
    \toprule
    \multirow{3}{*}{conditioning}       & \multirow{3}{*}{inversion} & \multicolumn{4}{c|}{\textbf{reconstruction}}                             & \multicolumn{4}{c}{\textbf{interpolation}}                               \\ \cmidrule(l){3-10} 
                                        &                            & \multicolumn{2}{c}{Nsynth (val set)} & \multicolumn{2}{c|}{Good-sounds}  & \multicolumn{2}{c}{Nsynth (val set)} & \multicolumn{2}{c}{Good-sounds}   \\ \cmidrule(l){3-10} 
                                        &                            & MSE$\downarrow$   & Pitch$\uparrow$  & MSE$\downarrow$ & Pitch$\uparrow$ & FID$\downarrow$   & Pitch$\uparrow$  & FID$\downarrow$ & Pitch$\uparrow$ \\ \midrule
    \multirow{2}{*}{pitch}              & enc.                       & 6.53              & 0.669            & 8.23            & 0.384           & 1451.7            & 0.296            & 2298.5          & 0.251           \\
                                        & enc. + opt.                & 6.34              & 0.629            & 8.44            & 0.126           & 1183.0            & 0.314            & 2292.4          & 0.214           \\
    \multirow{2}{*}{pitch + instrument} & enc.                       & 4.32              & 0.793            & 5.09            & 0.655           & 709.8             & 0.594            & 679.3           & 0.585           \\
                                        & enc. + opt.                & 3.49              & 0.778            & \textbf{3.12}   & 0.167           & 601.2             & 0.534            & 610.0           & 0.442           \\ \midrule
    \textbf{pitch + instance (ours)}    & -                          & \textbf{1.79}     & \textbf{0.904}   & 3.28            & \textbf{0.944}  & \textbf{252.2}    & \textbf{0.883}   & \textbf{477.4}  & \textbf{0.883}  \\ \bottomrule
  \end{tabular}
\end{table*}

\begin{figure}[t]
  \centering
     \includegraphics[height=4.0cm]{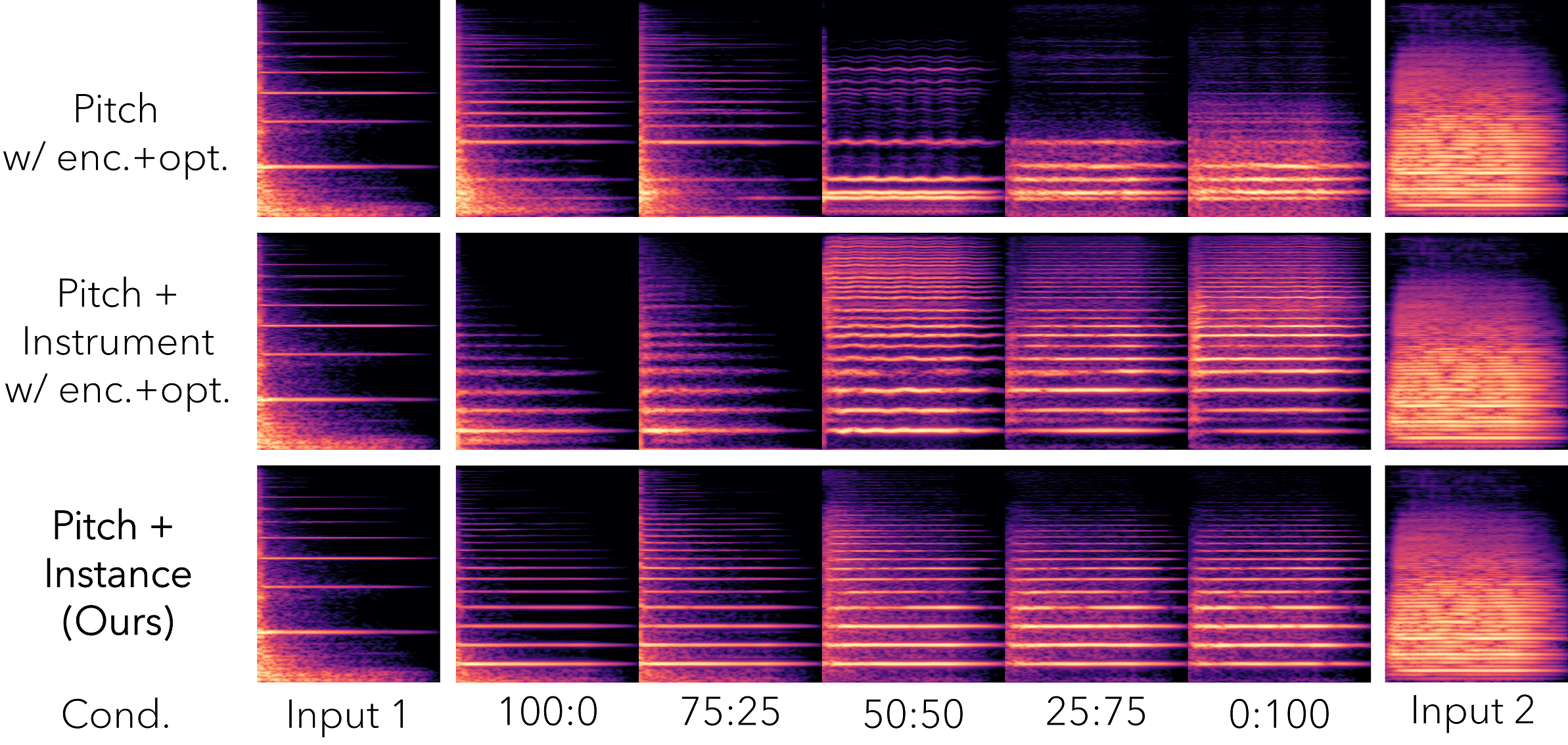}
     \vspace*{-4mm}
     \caption{Examples of interpolation in the latent space.\cref{1}}
     \label{fig:examples_of_reconstruction_and_interpolation}
\end{figure}

Next, we evaluated the editability of the input sounds.
To evaluate the faithfulness of reconstruction, we measured the mean squared error (MSE) and pitch accuracy of the synthesized samples.
MSE was computed on the feature space defined by the FID feature extractor.
To evaluate the ability of exploration in the latent space, we randomly chose two inputs from the dataset and interpolated the corresponding latent variables using a ratio sampled from uniform distribution $\mathcal{U}(0, 1)$ to generate interpolated samples.
We computed the FID between the input and interpolated samples.
We randomly chose a conditioning pitch from the 88 pitches.

The baseline models need to invert the inputs into the latent space for editing.
Typical approaches to GAN inversion can be categorized into learning-based, optimization-based, and a hybrid of the two \cite{xia2022gan}.
In our experiments, we used learning-based and hybrid approaches, which we found work well.
We trained encoders $E_1(\mathbf{x})$ and $E_2(\mathbf{x})$ for the baselines with objectives $\min_{E_1} || \mathbf{x} - G_1(E_1(\mathbf{x}), \mathbf{p}(\mathbf{x})) ||_2^2 + \lambda_\mathbf{z} || E_1(\mathbf{x}) ||_2^2$ and $\min_{E_2} || \mathbf{x} - G_2(E_2(\mathbf{x}), \mathbf{p}(\mathbf{x}), \mathbf{c}(\mathbf{x})) ||_2^2 + \lambda_\mathbf{z} || E_2(\mathbf{x}) ||_2^2$, respectively.
The second term is regularization for $\mathbf{z}$ to follow a standard normal distribution, which we found to be essential for the following optimization.
In the hybrid approach, we initialized latent variables with $\mathbf{z}=E_{\{1,2\}}(\mathbf{x})$ and minimized $L_2$ loss with respect to $\mathbf{z}$.

The middle portion of Table \ref{table:reconstruction_and_interpolation} shows that the baselines tended to fail in reconstructing the inputs and to sacrifice pitch accuracy, especially for Good-sounds, because they prioritize minimizing the reconstruction error, whereas GANStrument successfully reconstructed the inputs for both seen and unseen datasets.
The right side of Table \ref{table:reconstruction_and_interpolation} shows that the interpolated samples of the baselines produced a significant decrease in pitch accuracy, which suggests that the interpolated latents could deviate from the data manifold.
Our model, in contrast, had better FID and pitch accuracy, demonstrating that it can generate high fidelity samples with accurate pitch by exploring the latent space.

Fig. \ref{fig:examples_of_reconstruction_and_interpolation} shows qualitative examples of the interpolation in the latent space.
Their inputs were keyboard and brass sounds of the NSynth dataset and noise vectors $\mathbf{z}$ were fixed.
The baselines struggled with inverting the inputs and completely failed to mix two sounds.
In contrast, GANStrument smoothly interpolated two timbres with accurate pitch.

\mysubsection{Ablation study}
\label{subsec:ablation_study}
\begin{table}[t]
  \vspace*{-2mm}
  \caption{Ablation study: feature extractor}
  \label{table:ablation}
  \centering
  \begin{tabular}{@{}ccccc@{}}
    \toprule
    \multirow{2}{*}{feature extractor} & \multicolumn{2}{c}{Nsynth (train set)} & \multicolumn{2}{c}{Nsynth (val set)}   \\ \cmidrule(l){2-5} 
                                       & FID$\downarrow$ & Pitch$\uparrow$ & FID$\downarrow$ & Pitch$\uparrow$ \\ \midrule
    w/o adv. training                  & 95.3            & 0.731                & \textbf{191.4}  & 0.757                \\
    w/ adv. training                   & \textbf{90.4}   & \textbf{0.834}       & 212.3           & \textbf{0.870}       \\ \bottomrule
  \end{tabular}
\end{table}

\begin{figure}[t]
  \centering
     \includegraphics[height=3.1cm]{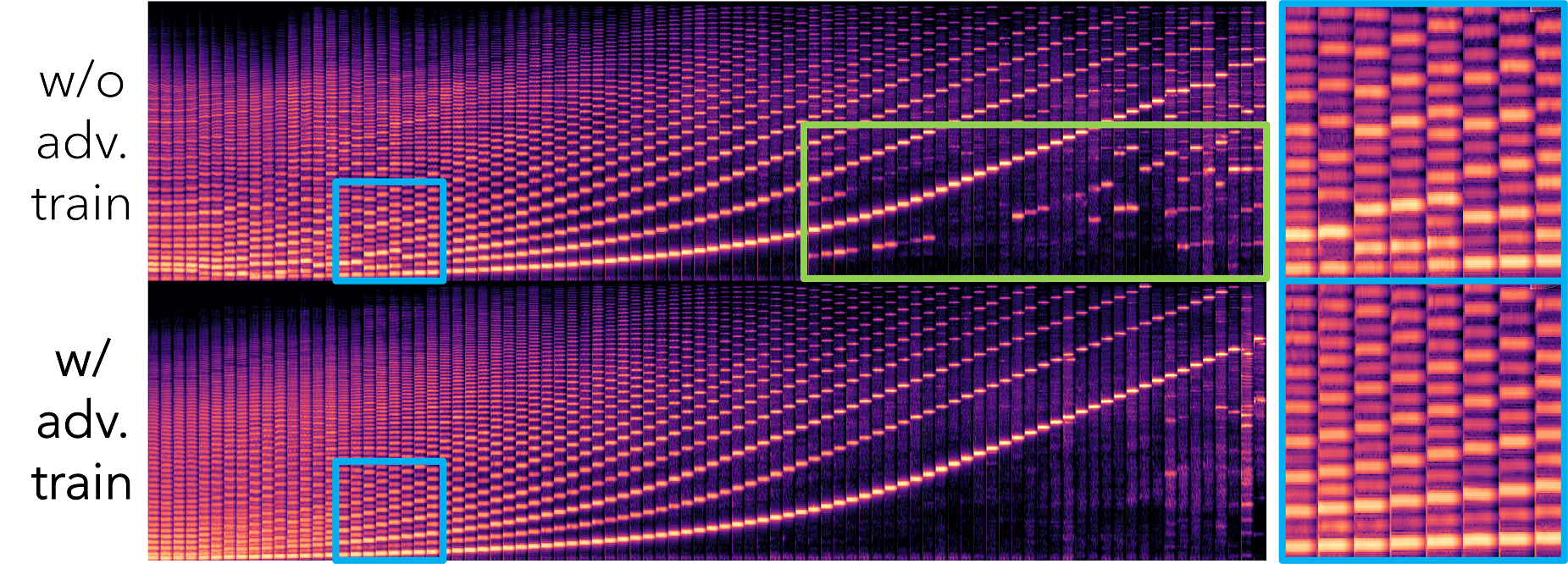}
     \vspace*{-4mm}
     \caption{Ablation study: difference between feature extractor without and with adversarial training.\cref{1}}
     \label{fig:ablation}
\end{figure}

Next, we conducted an ablation study to evaluate the effectiveness of the proposed pitch-invariant feature extractor.
For comparison, we trained an instrument identity classifier as a feature extractor $f_\phi$ using only the first term of Eq. \eqref{eq:feature_extractor_loss_fisrt}.
Table \ref{table:ablation} shows that our approach significantly improved pitch accuracy.
Fig. \ref{fig:ablation} shows mel-spectrograms of 88 pitches generated with the input of a saxophone sound of the Good-sounds dataset and a fixed noise $\mathbf{z}$.
The feature extractor without adversarial training produced inaccurate pitches, especially in higher tones as shown in the green box, as well as timbre inconsistency, as shown in the blue box.
The pitch-invariant feature extractor, in contrast, produced stable pitches with timbre consistency.

\mysubsection{Non-instrument sound inputs}
\begin{figure}[t]
  \centering
     \includegraphics[height=3.0cm]{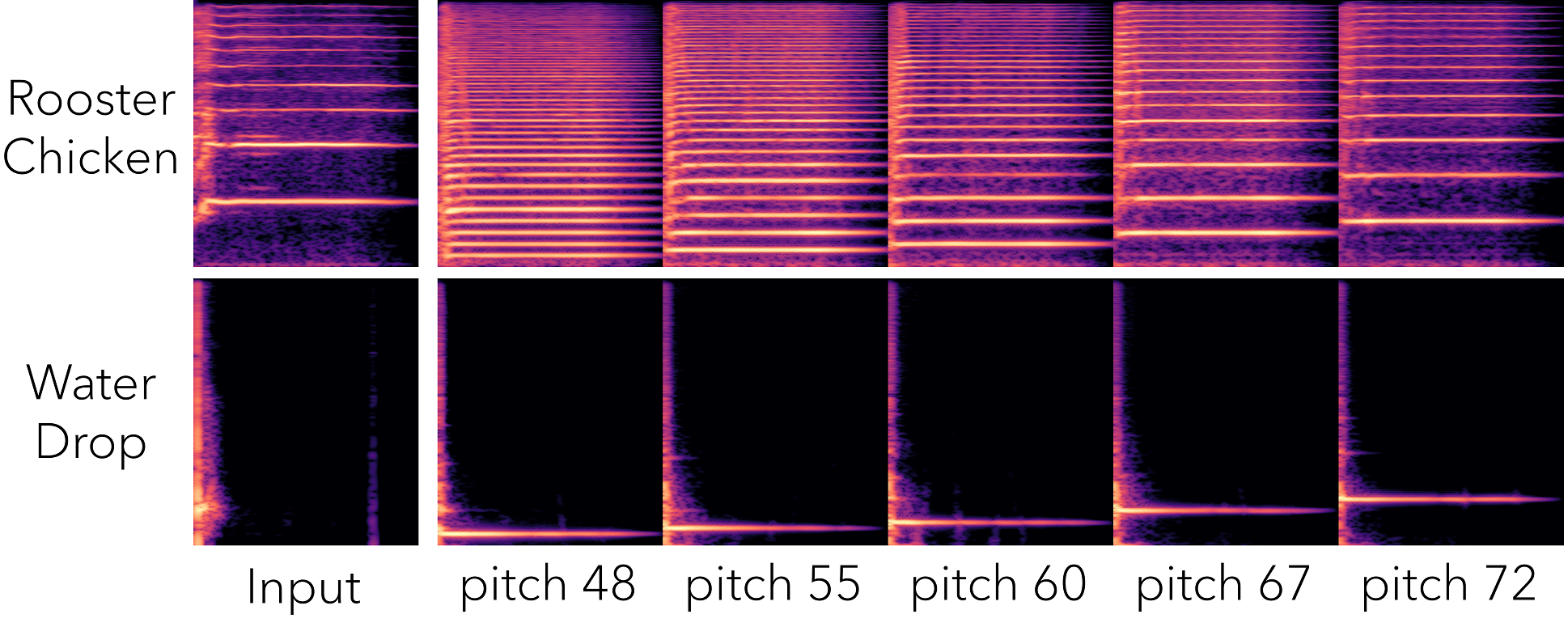}
     \vspace*{-4mm}
     \caption{Examples of non-instrument sound inputs.\cref{1}}
     \label{fig:generated_samples}
\end{figure}
Fig. \ref{fig:generated_samples} shows qualitative results with the input of non-instrument sounds such as a rooster and dropping water.
The synthesized sounds reflected the input timbres and produced stable pitch like musical instruments.
These results demonstrate that GANStrument has generalization ability to non-instrument sounds to some extent and is able to exploit a variety of sound materials to design the timbre as the traditional samplers do.
Note that the additional examples are available online\cref{1}.

\mysubsection{Timing}
Finally, we measured the generation timing on a middle-range CPU (Intel Core i7-7800X, 3.50 GHz).
The total time was 1.62 s, where the inference of the feature extractor $f_\phi$ and generator $G$ took 0.31 and 0.35 s, respectively, and mel-to-linear scale conversion 0.60 s, Griffin-Lim algorithm 0.36 s.
These results show that the improved mel-to-linear scale conversion described in Sec. \ref{subsec:audio_iversion} plays an important role in interactive generation.

\mysection{Related Work}
\label{sec:realted_work}

NSynth \cite{engel2017neural} uses a WaveNet \cite{oord2016wavenet}-based autoencoder to directly synthesize the waveforms of instrument sounds.
While it is capable of inference with a trained encoder, autoregressive sampling makes generation slow and prone to artifacts.
GANSynth \cite{engel2018gansynth} improves generation speed and quality by using an image synthesis model and a spectrogram with phase information.
However, it does not accept inputs, making it difficult to explore desired timbre in a complex latent space.
Luo et al. \cite{DBLP:conf/ismir/LuoAH19} proposed disentangling timbre and pitch using a Gaussian mixture VAE, but the simple architecture and autoencoder-based training make audio quality insufficient.

DDSP \cite{engel2020ddsp} and its subsequent work \cite{shan2022differentiable} achieve fast and interpretable generation by incorporating additive synthesis and wavetable synthesis into autoencoders.
However, their inputs should have clear pitch and the generated timbre is basically limited to a combination of integral multiples of the fundamental frequency \cite{engel2020ddsp}.
Leveraging the domain knowledge like these studies is complementary to our work and left for future work to further improve our model.

\mysection{Conclusion}
\label{sec:conclusion}
Our novel neural synthesizer, GANStrument, generates pitched instrument sounds reflecting one-shot input timbre within an interactive time.
It incorporates two key features: 1) instance conditioning, resulting in better generation quality and generalization ability to various inputs and 2) pitch-invariant feature extraction based on adversarial training, resulting in significantly improved pitch accuracy and timbre consistency.
Experimental results demonstrated the effectiveness of this approach.
We believe that GANStrument will enable users to generate novel instrument sounds as well as freely explore the desired timbre by utilizing a variety of sound materials.

%\vfill\pagebreak
\clearpage
\bibliographystyle{IEEEbib}
\bibliography{refs}

\end{document}